\documentclass{aipproc}
\usepackage{subfig}
\usepackage{natbib}
\usepackage{graphics}
\layoutstyle{8x11single}

\begin{document}
\title[]{Multi-hadron states in Lattice QCD spectroscopy}

\author{J. Foley}{address={Dept. of Physics, Carnegie Mellon University, Pittsburgh, PA 15213, USA}}
\author{J. Bulava}{address={NIC, DESY, Platanenallee 6, 15738 Zeuthen, Germany}}
\author{K.J. Juge}{address={Dept. of Physics, University of the Pacific, Stockton, CA 95211, USA}}
\author{C. Morningstar}{address={Dept. of Physics, Carnegie Mellon University, Pittsburgh, PA 15213, USA}}
\author{M. Peardon}{address={School of Mathematics, Trinity College, Dublin 2, Ireland}}
\author{C.H. Wong}{address={Dept. of Physics, Carnegie Mellon University, Pittsburgh, PA 15213, USA}}

\begin{abstract}
The ability to reliably measure the energy of an excited hadron in 
Lattice QCD simulations hinges on the accurate determination of 
all lower-lying energies in the same symmetry channel. These 
include not only single-particle energies, but also the energies 
of multi-hadron states. 
This talk deals with the determination of 
multi-hadron energies in Lattice QCD. The group-theoretical derivation 
of lattice interpolating operators that couple optimally to multi-hadron 
states is described. We briefly discuss recent algorithmic developments 
which allow for the efficient implementation of these operators 
in software, and present numerical results from the Hadron Spectrum Collaboration.
\end{abstract}

\maketitle
A stated aim of the Hadron Spectrum Collaboration is the 
determination of the low-lying hadron spectra across all flavor sectors 
for pion masses approaching the physical value. The ability to reliably 
extract a hadronic energy level from a lattice QCD simulation hinges on the 
accurate determination of all lower-lying energies in the same symmetry channel. 
Thus, a critical component of a spectroscopy calculation is the 
identification of a 
large and diverse set of interpolating operators that couple to all low-lying 
states.
The stationary-state energies accessible to Monte Carlo studies 
receive contributions not only from single-particle states, but also from multi-hadron
excitations.
Considerable effort has already gone into the identification of 
an optimal set of single-particle interpolators~\cite{Basak:2005aq,Dudek:2009qf},
and initial scans of the baryon, isovector meson, and kaon sectors have 
been promising~\cite{Morningstar:2010ae}. 
However, a complete analysis of the spectrum must include interpolating operators 
specifically designed to couple to multi-hadrons. 
The Wick contraction of multi-hadron operators can produce diagrams which contain 
quark lines that begin and end on a single time slice. In studies which use only 
single-particle operators, such diagrams are confined to the flavor-singlet meson 
channels. Until recently, precision measurements of the contributions from these 
diagrams to lattice correlation functions were prohibitively expensive.
Fortunately, due to recent advances in lattice algorithms~\cite{Morningstar:2010ae,Peardon:2009gh}, 
it 
is now feasible to accurately compute the contributions from these diagrams, 
 even on larger lattice volumes, and a comprehensive study of the spectrum including 
 the light isosinglet meson sectors and multi-particle states appears 
 within reach.

\subsection*{Group-theoretical construction of multi-hadron operators}
The multi-hadron operators in question are formed by combining 
single-particle operators (e.g. fermion bilinears in the meson sector) to form composite interpolators, which 
are expected to couple well to multi-particle states. 
To facilitate the extraction and identification of energies, 
interpolating operators are constructed to transform irreducibly under the 
lattice symmetry group. As in the continuum, operators project onto 
definite flavor sectors and meson interpolators have definite 
G-parity\footnote{Our simulations use a Wilson-type quark action 
  and are performed in the isospin limit: $m_{u} = m_{d}$.}
The lattice space group is the semi-direct product of the 
group of spatial lattice translations and the cubic point group $O_{h}$.
Irreducible representations (irreps) of the space group are 
conventionally labeled by a momentum vector $\mathbf{k}$ 
and a second 
label denoting irreps of the little group of $\mathbf{k}$.
In our numerical studies, we enforce periodic boundary conditions in the 
spatial directions. Hence, the components of the momentum vector $\mathbf{k}$ 
are quantized in units of $2 \pi/L_{s}$, where $L_{s}$ is the spatial 
extent of the lattice.

Hadron rest energies can be classified according to the $\mathbf{k} = \mathbf{0}$ 
space-group irreps. The corresponding little group is $O_{h}$,
which has ten single-valued irreps.
Following Mulliken convention, these are labeled 
$A_{1 g/u}$, $A_{2 g/u}$, $E_{g/u}$, $T_{1 g/u}$, $T_{2 g/u}$, 
 where the subscripts $g$ and $u$ denote even- (gerade) and odd- (ungerade) 
 representations respectively. Baryons at rest are classified according 
 to the fermionic, or double-valued,  irreps of $O_{h}$. 
These are obtained from the single-valued irreps of the double cubic group $O_{h}^{D}$. 
The double group has sixteen single-valued irreps.
Of these, ten are identical to the single-valued 
 irreps of $O_{h}$, while the remaining six representations, $G_{1 g/u}$, $G_{2 g/u}$, $H_{g/u}$, form  double-valued irreps of $O_{h}$.
 The simplest accessible multi-hadrons are made up of two
single-particle states at rest. However, on large spatial lattice volumes, most of the 
low-lying multi-hadron states are expected to consist of hadrons with non-zero relative momenta.
To date, we have analyzed the finite-momentum space-group irreps needed to 
construct operators for the lowest-lying multi-hadron states.
These include space-group irreps that correspond to momenta directed along lattice axes, such as $\mathbf{k} = (0,0,k)$, as 
well as momenta in planar-diagonal and cubic diagonal directions, for example, $\mathbf{k} =  \left( 0,k, k \right)$ and 
$\mathbf{k} = (k,k,k)$. The corresponding little groups are the point 
groups $C_{4 v}$, $C_{2 v}$, and $C_{3 v}$, respectively. 
$C_{4 v}$ has five single-valued irreducible representations 
and two fermionic irreps, $C_{3 v}$ has three single-valued irreps and three fermionic irreps,
 and $C_{2 v}$ has four single-valued irreducible representations and just one fermionic irrep. 
Table~\ref{table1} contains the decomposition of the single-valued irreps of $O_{h}$ subduced onto $C_{4 v}$ 
into irreps of $C_{4 v}$. 
 $A_{1}$, $A_{2}$, $B_{1}$ and $B_{2}$ are one-dimensional irreps, and 
 $E$ is two-dimensional.
The decompositions of the subduced representations show, for example, that finite-momentum operators
which transform according to the $A_{2}$ representation of $C_{4 v}$ project onto states with rest 
energies in the $A_{1 u}$, $E_{u}$ and $T_{1 g}$ irreps of $O_{h}$.
Space-group irreps for the lattice momentum vector 
$\mathbf{k}$ can be induced from the unitary irreps of the little 
group of $\mathbf{k}$.
The dimensions of these space-group representations are equal to the 
dimension of the star of
$\mathbf{k}$ (the set of distinct momentum vectors obtained by applying 
$O_{h}$ to $\mathbf{k}$) times the dimensions of the
little group irreps. Hence, the space-group irrep with momentum 
$(0,0,k)$ induced from the $A_{2}$ irrep of $C_{4 v}$ is six-dimensional.
A comprehensive discussion of the group theory for finite-momentum and multi-particle states can be found in Refs.~\cite{Moore:2005dw,Moore:2006ng}, 
whose results for the irreps which arise in this study we have independently verified.

\begin{table}
\begin{tabular}{l  l  l}
\hline
$ A_{1 g} \rightarrow  A_{1}  $ & & $~T_{1 g} \rightarrow A_{2} \oplus  E $ \\
$ A_{1 u} \rightarrow  A_{2}  $ & & $~T_{1 u} \rightarrow A_{1} \oplus  E $ \\
$ A_{2 g} \rightarrow  B_{1}  $ & & $~T_{2 g} \rightarrow B_{2} \oplus  E $ \\
$ A_{2 u} \rightarrow  B_{2}  $ & & $~T_{2 u} \rightarrow B_{1} \oplus  E $ \\
  $E_{g}~\rightarrow  A_{1} \oplus B_{1}   $ & &  \\
  $E_{u}~\rightarrow  A_{2} \oplus B_{2}   $ & & \\
  \hline
\end{tabular}
\caption{Decomposition of the single-valued irreps of $O_{h}$ 
  subduced onto $C_{4 v}$ into irreps of $C_{4 v}$.}
\label{table1}
\end{table}
To obtain a complete set of basis operators for the 
space-group irrep characterized by the lattice momentum $\mathbf{k}$ 
and the little group irrep $\Lambda$, we first identify a set of 
operators with definite momentum $\mathbf{k}$ that transform 
amongst themselves according to 
the irrep $\Lambda$ under the little group of $\mathbf{k}$.  
These form a subset of operators in the basis of the space-group irrep.
The other basis operators are obtained by applying rotations to this initial operator set.
To identify irreducible single-particle operators for the 
lattice little groups, we have adapted 
the operator construction algorithm described in Ref.~\cite{Basak:2005aq}. 
Irreducible operators are formed from  linear 
superpositions of more basic gauge-invariant operators, termed 
elemental operators.
The general expression for the single-particle elemental operators 
from which irreducible meson annihilation operators are constructed is 
\begin{eqnarray}
\phi^{A B}_{\alpha \beta,ijk} \left(t, \mathbf{k} \right)
&=&
\sum_{\mathbf{x}} e^{-i \mathbf{k} \cdot \mathbf{x} }
\left( \tilde{\chi}^{A}_{\alpha} \tilde{D}_{i}^{(p) \dagger} \right)
\left( \mathbf{x}, t \right)
  \left( \tilde{D}_{j}^{(p)} \tilde{D}_{k}^{(p)} \tilde{\psi}^{B}_{\beta} \right)
    \left( \mathbf{x}, t \right),
    \label{elemental}
\end{eqnarray}
where $\tilde{\chi} \left(\mathbf{x}, t\right)$ and $\tilde{\psi} \left( \mathbf{x}, t \right)$ 
represent smeared lattice quark fields with flavor indices $A$ and $B$ and spin indices $\alpha$ and $\beta$. 
$\tilde{D}^{(p)}_{i}$ is a $p$-link quark-field displacement operator: $\tilde{D}^{(p)}_{i} \rho \left( \mathbf{x}, t \right) 
  = \tilde{U}_{i} \left( \mathbf{x}, t \right) ... \tilde{U}_{i} \left( \mathbf{x} + (p-1) \hat{i}, t \right) \rho \left( \mathbf{x}+ p \hat{i}, t \right) $,
  for $i = \pm 1, \pm 2, \pm 3$,
  where $\tilde{U}$ are stout-smeared link variables.
  Our labeling convention also permits $i=0$, where $\tilde{D}_{0}^{(p)}$ is defined to be 
  a zero-displacement operator: $\tilde{D}_{0}^{(p)} \rho \left( \mathbf{x}, t \right) = \rho \left( \mathbf{x}, t \right)$.
A set of elemental operators which transform amongst themselves under the action of the little group 
(or, in the case of baryon operators, the corresponding double group)
generates a representation of the little group, $W$, which is in general reducible.
Given the unitary irrep matrices  $\Gamma^{\Lambda}$, one can define a projection matrix
\begin{eqnarray}
P^{\Lambda \lambda} = \frac{d_{\lambda}} {N_{D}} \sum_{g_{D}} \Gamma^{\Lambda}_{\lambda \lambda} \left( g_{D} \right) W \left( g_{D} \right),
  \label{projector}
\end{eqnarray}
where $d_{\Lambda}$ is the dimension of the irrep $\Lambda$, $g_{D}$ denotes an element of the double cover of the little group, and $N_{D}$ is the 
order of the double group. Applied to the elementals, this projection matrix returns a set of operators that transform 
according to the row $\lambda$ of the irrep $\Lambda$, provided that irrep appears in the decomposition of the  representation $W$; 
otherwise, $P^{\Lambda \lambda}$ is a null matrix. 
Note that the normalizations and phases of the projected operators are not known a priori. To get a consistent set of 
basis operators, we use projection matrices to generate a set of operators for the $\lambda=1$ row of the irrep only,
and employ the orthogonality relations for unitary irreps of finite groups to deduce the corresponding operators for other 
rows in the representation.
Bases of multi-hadron operators consist of  products of the single-particle operator sets.
The subsets of composite operators with zero net linear momentum generate (reducible) representations of 
$O_{h}$. In analogy to Eq.~\ref{projector}, we form projection matrices to get the required irreducible 
multi-hadron interpolating operators.

\begin{figure}
\resizebox{1.0\textwidth}{!}{\includegraphics*{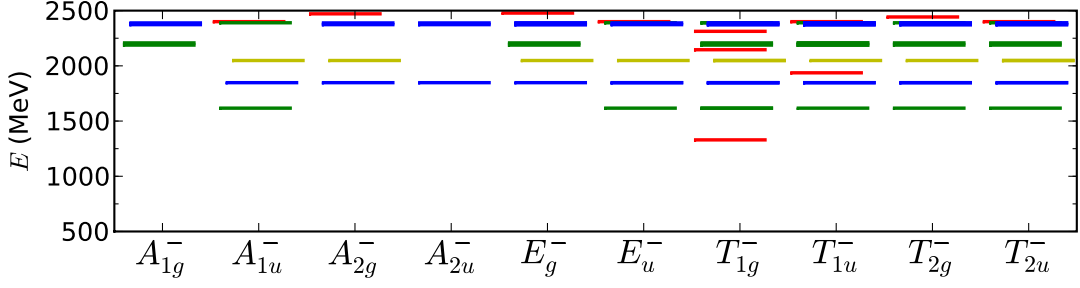}}
\caption{The low-energy spectrum of two non-interacting isovector mesons. 
  The energies are given by $ E = \sqrt{m_{1}^{2} + |\mathbf{k}|^{2}} + 
   \sqrt{m_{2}^{2} + |\mathbf{k}|^{2}}$, where $m_{1(2)}$ are measured meson rest energies.
  We show results for composite states with negative G-parity only (indicated by the 
      superscript on the irrep labels), and do not include error estimates.
  This spectrum was deduced from 
    measurements performed using single-particle operators at a 
    pion mas of $380~\rm{MeV}$ and a spatial volume of approximately $(1.9~\rm{fm})^{3}$.
    In each symmetry channel, a slight horizontal offset distinguishes energies
    with different relative momenta.
    The figure suggests that, even at these run-parameter values, multi-hadron states form a significant component 
    of the accessible hadron spectrum.}
\label{fig2}
\end{figure}

Having performed a first scan of the spectrum using single-particle operators, we get a rough estimate 
for the number and types of
multi-hadron operators required in each symmetry channel from the spectrum of states consisting 
of multiple non-interacting hadrons. 
Figure~\ref{fig2} shows results for systems of two non-interacting
isovector mesons with zero total momentum. The meson rest masses were obtained from low-statistics measurements with a
$ 380~\rm{MeV}$ pion on a $(1.9~\rm{fm})^{3}$ spatial volume. The figure contains energy levels corresponding 
to two mesons at rest, as well as states consisting of mesons with the minimum allowed lattice momenta along axes, 
and in planar-diagonal and cubic-diagonal directions. The energies of the moving single-particle 
states were calculated assuming $E^{2} = M^{2} + |\mathbf{k}|^{2}$, which is approximately true for 
small lattice momenta.
While the results of this qualitative analysis must be treated with caution, it 
does provide a ball-park estimate for multi-hadron thresholds and a first approximation
for the density of multi-hadron states in each symmetry channel.

By varying the displacement combinations in Eq.~\ref{elemental},
a large number of operator bases can be generated.
However, these bases may include operators that have little 
overlap onto the states of interest, as well as sets of operators 
that all couple strongly to the same state. 
Hence, a rigorous pruning procedure has been applied to 
the zero-momentum single-particle operators to identify a manageable 
subset of operators that couple to different low-lying states.
We  expect that the best multi-hadron operators 
are formed by combining single-particle operators that couple 
well to the constituent hadrons. Therefore, we apply the same 
rigorous selection process to the finite-momentum single-particle operators. 
Operator pruning is implemented as follows: first, we compute 
Euclidean-time correlation functions for a sample operator in each 
basis set in low-statistics Monte Carlo runs. The interpolating 
operators that produce the largest statistical errors are identified 
and discarded. From the remaining operator set, smaller subsets 
are used to compute the normalized correlation matrices 
$\hat{C}_{i j} \left( t \right) = C_{i j} (t) / \sqrt{C_{i i} \left( t \right) 
  C_{j j} \left( t \right) }$, 
  with $C_{i j} \left( t \right) = \langle \mathcal{O}_{i} \left(t \right)
  \overline{\mathcal{O}}_{j} \left( 0 \right) \rangle$.
  The subscripts $i$ and $j$ distinguish different operators 
  transforming according to a single row of a particular lattice irrep.
  We then select a set of operators that has a well-conditioned correlator 
  matrix at small values of $t$. A condition number close to unity indicates 
  that the operators are almost orthogonal and therefore couple to different states.

\begin{figure}
\resizebox{.5\textwidth}{!}{\includegraphics{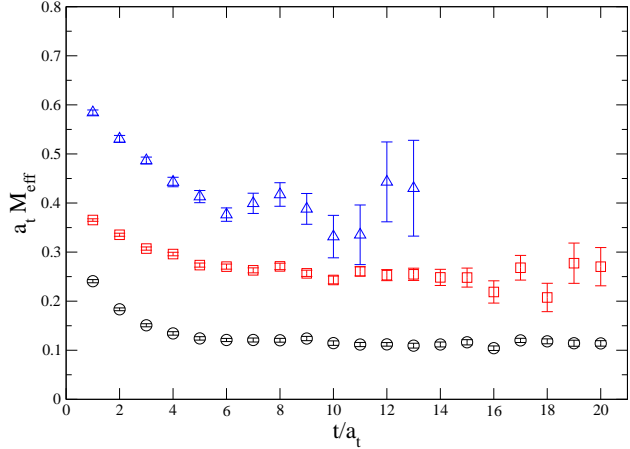}}
\caption{Effective masses for the three lowest-lying states 
  in a channel containing moving $\pi$ and $a$ mesons. 
    The meson momentum is 
    $L_{s} \mathbf{k} / \left( 2 \pi \right) = (0,0,1)$,
    and the figure shows results for the $A_{2}$ irrep of the
      little group 
      $C_{4 v}$. The resulting energy values are 
      consistent with measurements of the ground-state rest 
      masses in the $A_{1 u}$, $E_{u}$ and $T_{1 g}$ irreps of
      $O_{h}$.
      These measurements were performed on 93 configurations 
     using a set of six single-site and singly-displaced interpolating 
     operators which had been selected using the pruning procedure 
     outlined in the text. For clarity, a fourth level, which also 
     exhibits a reasonable plateau, has been omitted.
     The operators employed here will feature in the set of 
     single-particle operators used in the construction of composite, 
    multi-hadron operators.}
\label{fig1}
\end{figure}

In order to access the most important multi-hadron states on our target lattice volumes, 
we estimate that we require moving single-particle operators corresponding to the lowest 
two to four energy levels in most irreps. 
Our studies suggest that all the required operators can be formed from linear superpositions of local, or single-site, elementals 
and elemental operators containing a straight-line displacement in a single direction.
Figure~\ref{fig1} shows effective masses for single-particle isovector, negative G-parity meson operators with the 
momentum $L_{s} \mathbf{k} / \left( 2 \pi \right) = (0,0,1)$. These are results  for the $A_{2}$ irrep of 
$C_{4 v}$, which includes the energy of a pion in flight. 
The effective masses were extracted from a variational analysis of a $6 \times 6$ correlator matrix on 93
configurations.
There is a discernible plateau in each of the three lowest-lying effective masses, although due to low statistics the third level is somewhat noisy,
and we have verified that the fitted energies are consistent with ground-state rest-mass measurements in the $A_{1 u}$, $E_{u}$ and $T_{1 g}$ irreps.
The fourth level, which is not included in the figure, also exhibits 
an acceptable plateau. However, it overlaps the third level within errors, and has been 
omitted for the sake of clarity.

\subsection*{Computing multi-hadron correlators}
Finally, we turn to the recent algorithmic advances which facilitate the accurate evaluation of the required multi-hadron 
correlators on our target lattice volumes. We present only a brief outline
of progress in this area and refer the reader to Refs.~\cite{Morningstar:2010ae,Peardon:2009gh} for more detailed descriptions.

The first major development was the realization 
that one does not need to compute 
the quark-propagator components directly in order to evaluate hadron correlation functions;
instead, only the matrix elements of the 
propagator between eigenvectors of the quark-field smearing operator are required.
Since the smearing operator acts to suppress excited-state contamination, only 
a subset of smearing eigenvectors that preserves the important long-distance physics
needs to be included in the calculation. These ideas form the basis for a technique for computing 
correlation functions, known as distillation. 
A simple but effective implementation of this method involves defining the 
quark-field smearing operator to be a gauge-covariant laplacian with a hard cutoff 
imposed on high-momentum modes. This defines the Laplacian-Heaviside, or Laph, smearing scheme.

\begin{figure}[t] 
\begin{minipage}[b]{0.5\linewidth}
\vspace{-0.1cm}
\hspace{0.1cm}
\includegraphics*[scale=1.0]{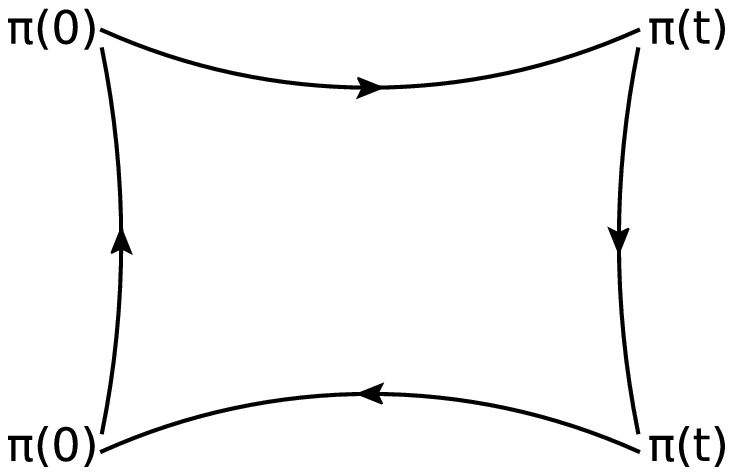}
\vspace*{0.8cm}
 \end{minipage}
 \hspace{0.1cm}
\begin{minipage}[b]{0.5\linewidth}
\includegraphics*[scale=0.3]{pipicorrelator}
 \end{minipage}
 \label{fig4}
 \caption{The left-hand side shows a quark-line diagram that arises in the 
   evaluation of a two-pion correlation function. 
     The plot on the right is the contribution of this diagram 
     to a correlator obtained using the stochastic Laph method described in the text.
     This test measurement was performed on 99 gauge-field configurations 
     generated at the three-flavor point on a $12^{3} \times 96$ lattice, and 
     the correlator was averaged over four source times.
     The operator used was designed to project onto two pions at rest.
     The dilution schemes for the quark lines connecting the source and sink 
     time slices and for the quark lines restricted to a single time slice, 
     which are given in parentheses, are explained in Ref.~\cite{Morningstar:2010ae}.
 }
\end{figure}

The distillation algorithm makes it feasible to compute the correlation functions
of interest on smaller lattice volumes. However, the number of smearing eigenvectors 
needed to capture the essential low-energy physics increases linearly with the 
spatial volume, reducing the efficacy of this method on larger lattices.
More recently, a new method for evaluating correlation functions, which combines distillation with a stochastic estimator, has 
been developed. The new algorithm  utilizes noise dilution~\cite{Foley:2005ac} to reduce 
the variance in correlation functions, and exact distillation is recovered in the 
maximal dilution limit.
In test studies, this method has consistently produced correlators with variances that are very close to the gauge-noise limit, 
   but at  considerably lower computational cost than exact distillation.
An example of results obtained for the box-diagram contribution to a two-pion correlation 
function using the stochastic Laph scheme is shown in figure~\ref{fig4}.

\subsection*{Conclusions}
Recent progress in lattice algorithms, combined with advances in computer 
infrastructure, has made it possible to incorporate explicit multi-hadron 
operators into our spectroscopy calculations. Our current focus is the construction 
of sets of multi-hadron operators for each lattice symmetry channel. 
To that end, we have put considerable effort into identifying good single-particle 
operators for hadrons with non-zero momenta. Pruning of the finite-momentum isovector 
meson operators is complete, and operator pruning in other meson and baryon sectors is underway.
We plan to begin measurements using the resulting operator set in the coming months.

This work was supported by the U.S. National Science Foundation under awards 
PHY-0510020, PHY-0653315, PHY-0704171 and through TeraGrid resources 
provided by the Pittsburgh SuperComputer Center, the Texas Advanced Computing 
Center, and the National Institute for Computational Sciences. MP is 
supported by Science Foundation Ireland under research grant 
07/RFP/PHYF168. 
We thank our colleagues within the Hadron Spectrum Collaboration.
Numerical calculations were performed using the Chroma software suite~\cite{Edwards:2004sx}. 

\bibliographystyle{unsrtnat}
\bibliography{hadron09}

\end{document}